# Predicting Long-term Urban Overheating and Their Mitigations from Nature Based Solutions Using Machine Learning and Field Measurements

Jiwei Zou[a,b], Lin Wang[a,*], Senwen Yang[b], Michael Lacasse[a], Liangzhu (Leon) Wang[b]

[a] Construction Research Centre, National Research Council Canada, 1200 Montreal Road, K1A 0R6, Ottawa, Ontario, Canada

[b] Centre for Zero Energy Building Studies, Department of Building, Civil and Environmental Engineering, Concordia University, Montreal, Canada

[*]Email: Lin.Wang@nrc-cnrc.gc.ca; leon.wang@concordia.ca

## Abstract

Urban overheating has become a global issue, exacerbated by climate change and that may lead to severe effects on both public health as well as urban sustainability. Traditional methods, such as the use of numerical simulations and field measurements, strive to accurately predict future urban overheating and the effects of urban greening since this is a challenging field given the many uncertainties in the input data. This study is intended to permit the prediction of the longevity and severity of future urban overheating events by integrating field measurements and machine learning models, focusing on the impact of urban greening under different global warming (GW) scenarios. Field measurements have been conducted during summer 2024 in an office campus at Ottawa, a city located in cold climate zone. Microclimate data were measured at four locations within the campus, the four locations have different types and coverage levels of urban greenings – large lawn area without trees (Lawn), parking lot without any greening (Parking), greenery area with sparsely distributed trees (Tree) and an area with 100% coverage of trees (Forest).  Models, such as Artificial Neural Networks (ANN), and Recurrent Neural Networks (RNN), and Long Short-Term Memory network models (LSTM) were trained on local microclimate data, with LSTM chosen for its superior performance predictions. Four Global Warming (GW) scenarios were considered to represent different Shared Socioeconomic Pathways (SSP) by 2050 and 2090. The results show that the UTCI at the "Parking" location increased from around 27 °C under GW1.0 to 31 °C under GW3.5. Besides, low health risk (UTCI > 26 °C) will be increased in all locations due to climate change impacts, regardless of urban greening conditions. However, the tree area like 'Tree' and 'Forest' are effective in eliminating the occurrence of extremely high-risk heat conditions (UTCI > 38.9 °C). The findings demonstrate that urban greening plays a crucial role in reducing severe thermal stress, thereby enhancing thermal comfort under future climate scenarios.

## 1. Introduction

Urban overheating has been a global issue in recent years, bringing severe impacts on public health and safety [1]. Historical heatwaves have claimed numerous lives, particularly among vulnerable populations located in urban areas. For instance, the 2003 European heatwave resulted in over 30,000 fatalities, and more recent heat events, such as the 2021 heatwave in the province of British Columbia, Canada, claimed around 500 lives [2-4]. Based on Zou *et al.* [3], average monthly overheating hours would be doubled by the mid-century and tripled, or indeed, quadrupled, by the end of the century. The increasing frequency and severity of extreme heat events have underscored the importance of understanding urban overheating as a means to protect public health, mitigate economic losses, and improve the overall quality of life in urban settings.

Urban overheating is expected to become even more intense and widespread under climate change, creating serious challenges for urban sustainability and livability [4-10]. Climate projections, as based on Shared Socioeconomic Pathways (SSP) scenarios, have been shown to indicate a significant increase in global temperature. For example, under the moderate SSP2-4.5 scenario, global temperatures are expected to rise by approximately 2.0°C above pre-industrial levels by 2050, whereas under the worst-case SSP5-8.5 scenario, temperatures could increase by as much as 4.4°C by 2090 [11]. These rising temperatures will affect thermal comfort of the public and increase health risks, particularly in densely populated cities having limited extents of urban greening [12-14]. Evaluations using future climate projections and methods for the selection of suitable reference years, have demonstrated that average monthly outdoor and indoor overheating hours could increase by up to 150% under Representative Concentration Pathways (RCP) 8.5 by the mid-term (2041-2060), and by up to nine (9) times by the end of the century (2081-2100) [3]. As well, a study by Zou *et al.* [12] indicates, under typical conditions, a shift from "Slight cold stress", as occurred in the 2010s, to "Extreme heat stress" in the 2090s, with "Extreme heat stress" becoming increasingly common under extremely warm conditions in a Montreal downtown area. Thus, understanding and predicting the impacts of these future climate scenarios on urban overheating is crucial for developing effective mitigation strategies, such as urban greening, to enhance thermal comfort and reduce health risks of the population residing in Urban settings.

There are two main approaches to assessing urban overheating: numerical simulations and field measurements. Numerical simulations, such as Computational Fluid Dynamics (CFD) [15], Weather Research and Forecasting models (WRF) [16], and the Global Environmental Multiscale Model (GEM) [17], have been widely used to evaluate urban microclimates and overheating. The CFD simulations can resolve the transfer of heat, mass, and momentum at fine spatial resolutions down to 1 m, thereby capturing details of urban microclimates, particularly around buildings or street canyons [12, 18-26]. On the other hand, WRF and GEM models are suitable for simulating urban overheating over larger spatial and temporal scales [27-29]. Whereas these simulation tools provide valuable spatial insights, they are computationally expensive and require realistic and sophisticated settings for boundary conditions. Thus, using numerical tools to predict future urban overheating can be challenging. This is particularly true when aiming to ensure accuracy and account for various complex physical processes, such as mitigation strategies involving urban greening. Field measurements, meanwhile, provide localized and highly accurate data but are limited by their spatial coverage and scalability [30-32]. Due to the sensor deployment and localized influences on the collected data, field measurements are often restricted to a few representative sites. They may not capture the diversity of conditions across an entire urban landscape [33-35]. Despite these constraints, their role in validating numerical simulations is crucial, ensuring that model predictions align with real-world conditions [18].

Machine learning (ML) models have recently drawn considerable attention due to their ability to bridge the gap between accuracy and scalability in assessing urban overheating [4, 32, 36]. Machine learning approaches, such as Random Forest (RF), Artificial Neural Networks (ANN), and Recurrent Neural Networks (RNN), have shown promising accuracy in predicting urban microclimate dynamics based on historical data [37-39]. These models can learn complex patterns from large climate datasets, enabling them to make accurate predictions without extensive computational resources and/or restrictions on boundary conditions. Machine learning models also have the ability to generalize across different urban settings, making them well-suited for capturing the spatial and temporal variability inherent in urban environments [40-42]. By training on field measurements and utilizing large-scale future climate data, machine learning offers a scalable solution for predicting future local urban overheating while maintaining an acceptable level of accuracy.

Urban greening has emerged as a promising mitigation strategy for reducing urban overheating and enhancing thermal comfort [9, 43, 44]. Urban vegetation, including the use of green roofs, green walls, and trees, plays a critical role in regulating temperatures in urban settings by providing shade, increasing evapotranspiration, and reducing surface temperatures [45-47]. However, the effectiveness of urban greening in mitigating urban overheating under future climate change conditions remains difficult to predicted, thus representing a crucial gap in current research. Due to the complexities involved in simulating the effects of urban greening across long-term periods using numerical models, existing studies have not adequately addressed this aspect [4, 48-50].

This study integrates field measurements and machine learning models to predict future urban overheating at local sites with limited measurement data under various urban greening settings. On-site field measurements were conducted on an office campus at the city of Ottawa during summer months at four locations with different urban greening settings, as illustrated in Section 2.1. Three machine learning methods are introduced in Section 2.2; Section 2.3 describes the future climate data that was used as input of the ML models for future urban microclimate prediction and Section 2.4 shows the thermal comfort index used for overheating evaluation. The results and discussions section include the description of field measurements in Section 3.1, the performances of the ML models in predicting local microclimate in Section 3.2, as well as the impacts of climate change and urban greening on urban overheating risks in Section 3.3.

# 2. Methodology

## 2.1 Field measurement on office campus

2.1.1 Study area and time period

Onsite measurements were performed inside the NRC campus, as shown in Fig. 1 (b), in the city of Ottawa (Fig. 1 a). Four different locations were selected for conducting onsite measurements, each having different urban features. Fig. 1 (c) shows the weather station placed in the middle of a grass area of land, indicated as the 'Lawn' site throughout this paper. this site is recognized as an open space with no surrounding trees or buildings within a 50-meter radius. Thr 'Lawn' site was selected for its relatively high exposure to wind speed and full solar radiation with moderate humidity and air temperature. Fig. 1 (d) shows the weather station at a parking lot within the NRC campus, indicated as the 'Parking' site. This measurement site occupies two parking spaces and is about 15 meters from the surrounding buildings. The 'Parking' site was selected for its relatively high wind speed and air temperatures, low humidity, and full exposure to solar radiation. Fig. 1 (e) shows the weather station located under a tree inside a small park at the NRC campus, indicated as the 'Tree' site. The weather station is about 2 meters away from the nearest tree, benefiting from the shading of the trees. The 'Tree' site was selected for its shaded solar radiation, moderate air temperature, humidity, and wind speed. Finally, Fig. 1 (f) shows the weather station placed within a forest inside the NRC campus, indicated as the 'Forest' site. The weather station was placed on a trail within the forest, surrounded by densely distributed canopy of trees approximately 20 meters in height. The 'Forest' site was selected for its relatively low solar radiation, wind speed, air temperature, and high relative humidity.

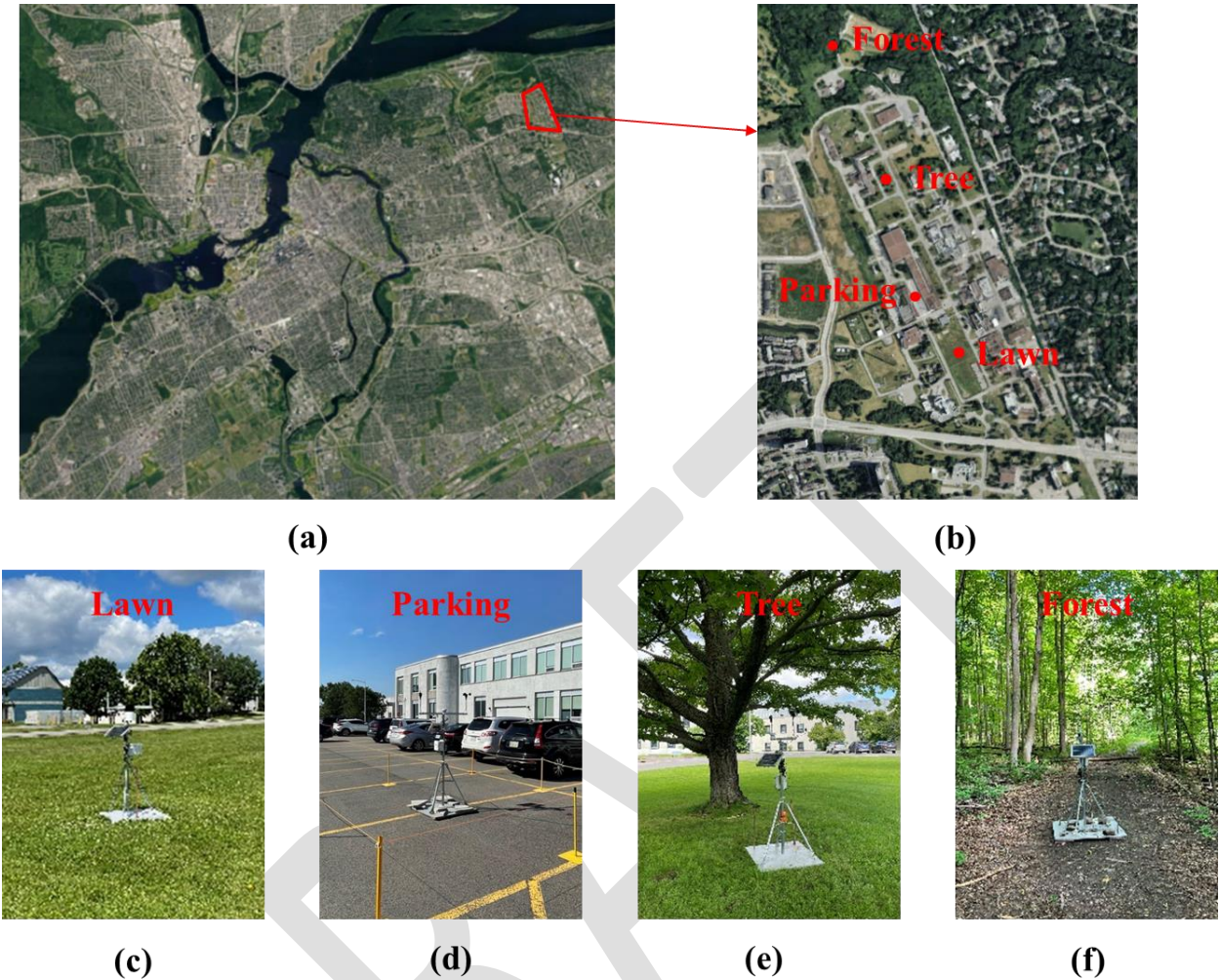

Fig. 1 Measuring locations of the study. (a) Map of Ottawa-Gatineau region (Red polygon represents the office campus). (b) Map of the campus. (c) Weather station on Lawn. (d) Weather station in Parking Plot. (e) Weather station under Tree. (f) Weather station in Forest.

The field measurements at the NRC campus were conducted from June 15$^{th}$ to September 14$^{th}$, 2024, covering three summer months. In Fig. 2 the daily air temperature, wind speed, relative humidity, and solar radiation are given as recorded at the Ottawa International Airport (YOW) for the same monthly period in 2024. The YOW was selected to collect climate data for Ottawa due to the availability of its long-term climate prediction. During these summer months, the daily air temperature ranged from approximately 8°C to 28°C, wind speed varied between 1 m/s and 6.8 m/s, relative humidity ranged from 52% to 100%, and solar radiation fluctuated between 0 W/m² and 400 W/m².

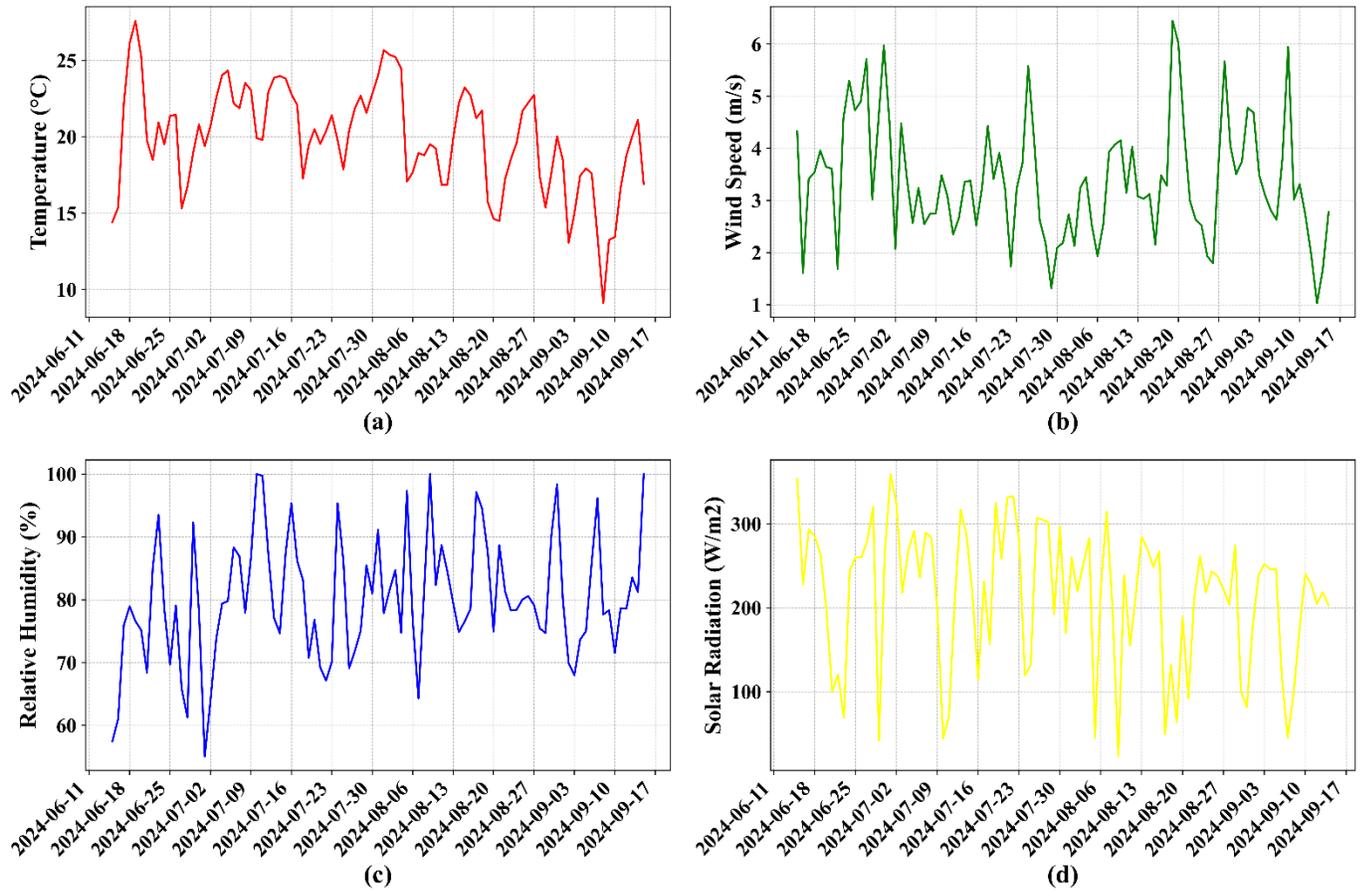

Fig. 2 Daily climatic variables reported at Ottawa international airport during 2024 summertime. (a) Daily Air Temperature. (b) Daily Wind Speed. (c) Daily Relative humidity. (d) Daily solar radiation.

2.1.2 Microclimate measuring system

Fig. 3 provides details of the weather station used in this study. All four weather stations were equipped with the same Air Temperature/Relative Humidity sensor. However, only the 'Tree' and 'Forest' stations were fitted with solar radiation sensors, as these two locations were shaded by trees. The 'Lawn' and 'Parking' stations were fully exposed to solar radiation throughout the day and therefore, the solar radiation sensor was not installed in those stations. The solar radiation data measured in the airport was used to fulfill the solar radiation data in these two locations, and calculate the thermal comfort indices for these two locations. Additionally, the 'Forest' station used an ultrasonic wind sensor to capture relatively low wind speeds, while the other three stations ('Lawn', 'Parking', and 'Tree') used a standard three-cup anemometer and wind vane, as shown in the figure.

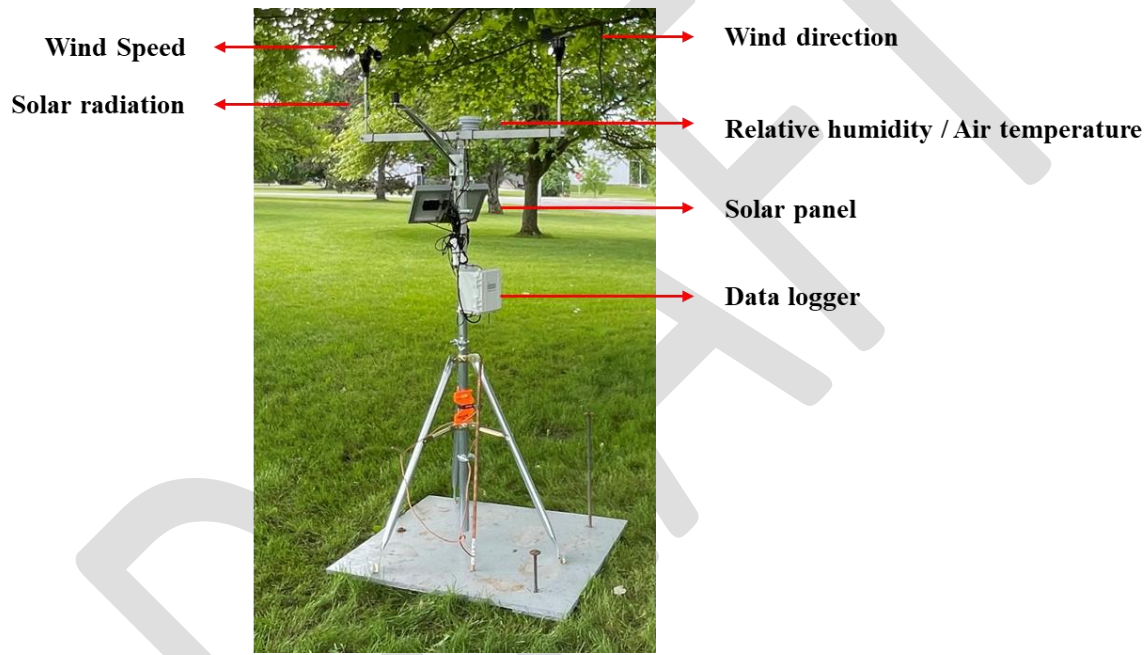

Fig. 3 Details of the weather station of 'Tree'

Table 1 summarizes the technical specifications of all the measurement instruments used in the study. The weather stations, positioned at a height of 1.8 m, recorded air temperature, relative humidity, wind speed, wind direction, and solar radiation. This height was chosen to represent conditions at the pedestrian level while minimizing the influence of activities such as walking and vehicle movement. The HOBO RX3000 Remote Monitoring Station was utilized for monitoring, recording, and transmitting climate data from each weather station. All climatic variables were recorded as one-minute averages and exported at a minute-level resolution. Solar panels were used to power the instruments at each weather station.

Table 1 Technical specifications of all the measurement instruments

| Sensor | Model | Parameter | Range | Resolution | Accuracy | Threshold |
|---|---|---|---|---|---|---|
| Temperature/RH Smart Sensor | S-THC-M008 | Temperature | -40 ~ 75 °C | 0.02 °C | ± 0.2 ~ 0.25 °C | - |
| | | Relative humidity | 0 – 100 % | 0.01 % | ± 2.5% | - |
| Wind Direction Smart Sensor | S-WDA-M003 | Wind direction | 0 ~ 355 degrees | 1.4 degrees | ± 5 degrees | 1 m/s |
| Wind Speed Smart Sensor | S-WSB-M003 | Wind speed | 0 ~ 76 m/s | 0.5 m/s | ± 1.1 m/s | 1 m/s |
| Silicon Pyranometer Smart Sensor | S-LIB-M003 | Solar Radiation | 0 ~ 1280 W/m$^2$ | 1.25 W/m$^2$ | ± 10 W/m$^2$ | - |
| Ultrasonic Wind Speed and Direction Smart Sensor | S-WCG-M003 | Wind Speed | 0 ~ 41 m/s | 0.4 m/s | ± 0.8 m/s | - |
| | | Wind Direction | 0 ~ 359 degrees | 1 degree | ± 4 degrees | - |

## 2.2 Machine learning methods -ANN, RNN, LSTM

This study is intended to provide an approach to predicting the future local microclimate of a selected area using machine learning methods and based on limited field measurement data. The general approach involves training machine learning models with three months of climate data collected from the local airport and, as well, four locations ('Forest', 'Lawn', 'Parking', and 'Tree') on the NRC campus. According to Yang *et al.* [51], the proposed approach would not increase the prediction accuracy if more than three months of training data was collected. The data used as input were collected from the Ottawa International Airport (YOW) and includes: hourly air temperature, relative humidity, wind direction, wind speed, and solar radiation. The output (target) variable for each model was measured by the hourly climatic variable (air temperature, wind speed, relative humidity, solar radiation) available at each measurement site. The structure of ANN, RNN, and LSTM models is shown in Fig. 4.

Before training the models, the data was preprocessed by removing any missing values and then normalized using the StandardScaler from sci-kit-learn [52]. A time-series approach was also adopted, where the LSTM and RNN models used a sliding window of 24 hours to capture temporal dependencies within the data [53, 54]. For model training and evaluation, the dataset was split into training and testing sets using an 80-20 ratio, where 80% of the data was used for model training, and the remaining 20% was reserved for testing, as based on the literature [55-59]. Model performance was evaluated using correlation coefficient scores (R²) and mean absolute error (MAE) to assess the accuracy and robustness of predictions on both training and test sets [37, 60].

Meanwhile, the accuracy of the model has been tested by calculating the MAE in Equation (1) and the correlation coefficient ($R^2$) in Equation (2).

$$MAE = \frac{\sum_{i=1}^{n}|y_i - \dot{y}_i|}{n} \quad (1)$$

$$R^2 = \frac{\sum_{i=1}^{n}(y_i - \dot{y}_i)}{\sum_{i=1}^{n}(y_i - \bar{y}_i)} \quad (2)$$

Where $y_i$ is local measured climatic data inside test set, $\dot{y}_i$ is local climatic data predicted by machine learning methods, $n$ is the number of climatic data of testing dataset, $\bar{y}_i$ is the average value of $y_i$.

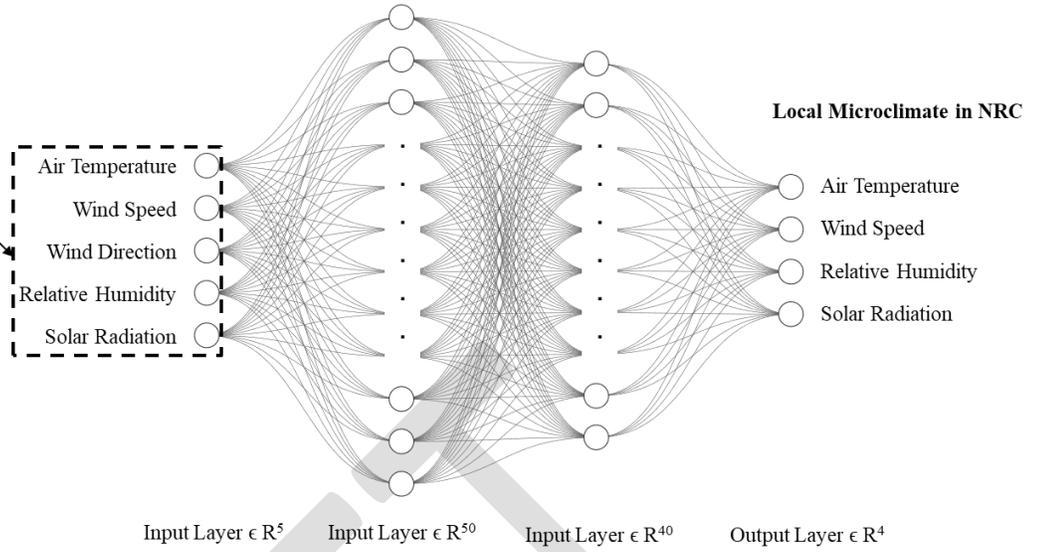

(a)

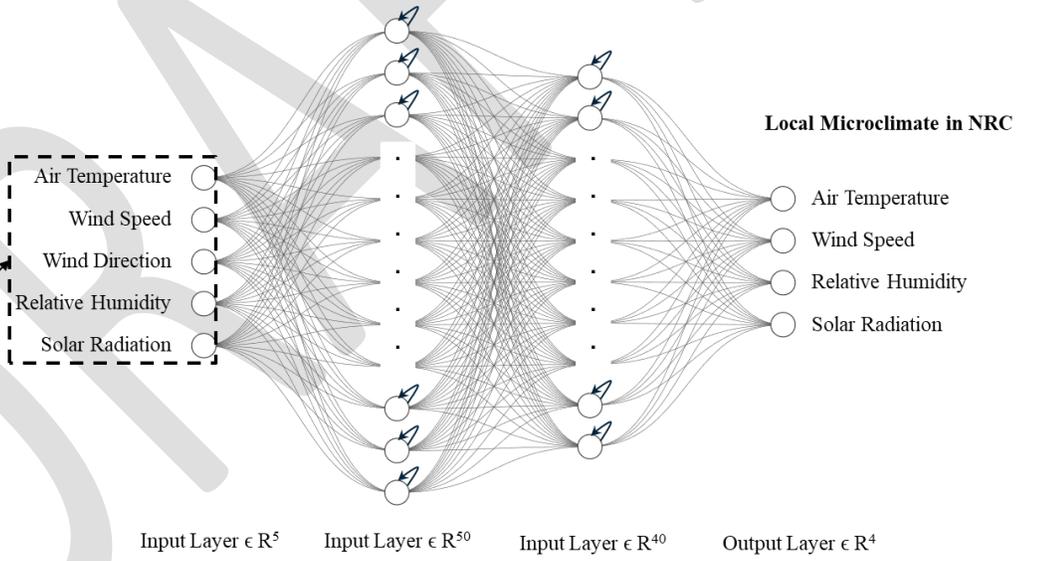

(b)

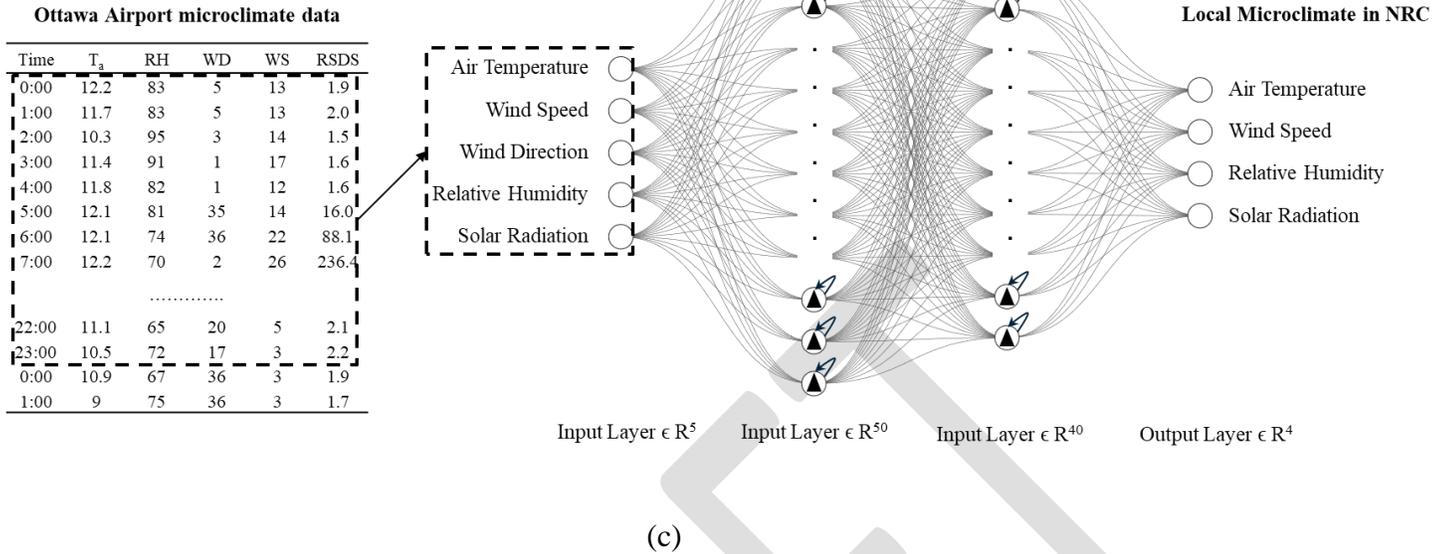

(c)

Fig. 4 Structure of (a) ANN and (b) RNN and (c) LSTM model

The Artificial Neural Network (ANN) model used in this study is a fully connected feed-forward network that consists of three layers: an input layer, two hidden layers, and an output layer, as shown in Fig. 4 (a). The hidden layers have 50 and 40 neurons, respectively, and use the ReLU activation function, whereas the output layer has a linear activation function. The ANN model was chosen for its ability to approximate complex non-linear relationships between the input features and the target variable. The ANN cells do not have any memory of previous inputs and are not designed to handle sequential data, which limits their ability to capture temporal relationships. One advantage of the ANN is its relatively simple architecture, allowing fast training and easy interpretability. However, it may not be as effective as other models for capturing temporal dependencies.

The Recurrent Neural Network (RNN) model, specifically a SimpleRNN, was also employed to capture the sequential nature of the climate data, as shown in Fig. 4 (b). The RNN model included two recurrent layers with 50 and 40 neurons, respectively, along with dropout layers with a rate of 20% to reduce overfitting. RNNs have neural cells with recurrent connections, which means they can retain information from previous time steps through feedback loops. This allows RNNs to model sequential data by maintaining an internal state, but they often struggle with retaining information over long sequences due to issues such as vanishing gradients. Whereas the RNN model is computationally less intensive and faster to train, it may not capture the temporal relationships as effectively as Long Short-Term Memory (LSTM) network models.

The Long Short-Term Memory (LSTM) network was used to handle the temporal dependencies present in the climate data, as shown in Fig. 4 (c). The LSTM model includes two LSTM layers with 50 and 40 neurons, respectively, followed by dropout layers with a rate of 20% to prevent overfitting. LSTM network models build upon RNNs by incorporating a more complex cell structure with gating mechanisms—input, forget, and output gates. These gates help regulate the flow of information, allowing LSTM cells to effectively learn long-term dependencies, making them better suited for capturing temporal patterns over long sequences compared to standard RNNs. The LSTM network is well-suited for time-series data because of its ability to maintain long-term dependencies through its memory cell structure. This makes it particularly advantageous for predicting microclimatic conditions, as it can capture the relationship between current and past observations. Despite being computationally more intensive compared to traditional feed-forward networks like ANN and even RNNs, the LSTM model was essential for capturing the temporal patterns in the microclimate data, which ultimately improved prediction accuracy.

## 2.3 Future climate data of Ottawa

The Shared Socioeconomic Pathways (SSP) provide a range of future global warming scenarios, each representing different socioeconomic pathways and climate policies [59]. For this study, five

SSP scenarios were considered to examine the impact of global warming levels on local thermal comfort performance by both 2050 and 2090:

- **SSP1-1.9**, representing the most optimistic scenario, aims to limit global warming to 1.6°C above pre-industrial levels by 2050 and remains below this threshold by 2090. This scenario requires rapid decarbonization and robust climate policies to mitigate emissions effectively.
- **SSP1-2.6** is another optimistic scenario where global warming reaches 1.7°C by 2050 compared to pre-industrial levels and stabilizes at 1.8°C by 2090.
- **SSP2-4.5** represents a scenario with moderate mitigation efforts. By 2050, global warming is expected to reach 2.0°C above pre-industrial levels, while by 2090, it could reach 2.7°C.
- **SSP3-7.0** is a more pessimistic scenario, reflecting higher emissions and limited mitigation efforts. In this scenario, warming reaches 2.1°C above pre-industrial levels by 2050.
- **SSP5-8.5** represents the worst-case scenario with very high greenhouse gas emissions and minimal climate action. Under this scenario, global warming is projected to reach 2.4°C by 2050 and 4.4°C by 2090 relative to pre-industrial levels.

Future climate datasets generated from CanRCM4 and bias corrected by Gaur and Lacasse [60] were used for simulations. In Gaur and Lacasse [60], the future climate data were prepared for seven global warming (GW) levels (GW0.5, GW1.0, GW1.5, GW2.0, GW2.5, GW3.0, and GW3.5) compared to a reference period from 1991 to 2021, each GW level includes 15 realizations (runs) of 31-year climate data, reflecting uncertainty of initial conditions in CanRCM4 modelling. To select appropriate GW levels that reflect SSP scenarios being considered over the next 25 and 65 years, the GW levels prepared by Gaur and Lacasse [60] were adjusted by approximately 0.6°C to convert the reference period to the pre-industrial period [59]. Four selected GW levels to represent SSP scenarios are presented in Table 2.

Table 2 GW scenarios and its corresponding SSP scenarios by 2050s and 2090s

| Selected GW levels to 1991-2021 | BY 2050 | | | BY 2090 | | |
|---|---|---|---|---|---|---|
| | SSP scenarios | GW to pre-industry | GW to 1986-2005 | SSP scenarios | GW to pre-industry | GW to 1986-2005 |
| GW1.0 | SSP1-1.9 SSP1-2.6 | 1.6~1.7 | 1.0~1.1 | SSP1-1.9 SSP1-2.6 | 1.4~1.8 | 0.8~1.2 |
| GW1.5 | SSP2-4.5 SSP3-7.0 | 2.0~2.1 | 1.4 | - | - | - |
| GW2.0 | SSP5-8.5 | 2.4 | 1.8 | SSP2-4.5 | 2.7 | 2.1 |
| GW3.5 | - | - | - | SSP5-8.5 | 4.4 | 3.8 |

**2.4 Thermal comfort index**

The Universal Thermal Climate Index (UTCI) was developed for assessing thermal comfort under dynamic air temperature and relatively high metabolic rate conditions, making it more suitable for evaluating outdoor thermal comfort than other prevailing thermal comfort index, such as Predictive Mean Vote (PMV), Physiological Equivalent Temperature (PET), and Standard Equivalent Temperature (SET). The thermal stress catalogue corresponding to UTCI (°C) values was listed in Table 3.

Table 3 Thermal stress category based on the value of UTCI.

| UTCI (°C) | Stress Category |
|---|---|
| 46 ≤ UTCI | Extreme heat stress |
| 38 ≤ UTCI < 46 | Very strong heat stress |
| 32 ≤ UTCI < 38 | Strong heat stress |
| 26 ≤ UTCI < 32 | Moderate heat stress |
| 9 ≤ UTCI < 26 | No thermal stress |
| 0 ≤ UTCI < 9 | Slight cold stress |
| -13 ≤ UTCI < 0 | Moderate cold stress |
| -27 ≤ UTCI < -13 | Strong cold stress |
| -40 ≤ UTCI < -27 | Very strong cold stress |
| UTCI < -40 | Extreme cold stress |

Based on its definition [61-63], UTCI is a function of air temperature ($T_a$), wind speed ($V_{w\_10}$), relative humidity ($RH$), and mean radiant temperature ($T_{mrt}$).

$$UTCI = f(T_a; V_{w\_10}; RH; T_{mrt}) \qquad (3)$$

In addition, UTCI requires the wind speed at the elevation of 10 m above the ground. Thus, in this study, the wind speed ($V_w$) from measurements is transformed into the wind speed at 10 m height through the power law equation:

$$\frac{V_w}{V_{w\_10}} = \left(\frac{h_{pd}}{h_{ref}}\right)^\alpha \qquad (4)$$

Here, $V_w$ is the wind speed measured at the pedestrian level height ($h_{pd}$) of 1.8 m. $V_{w\_10}$ is estimated wind speed at 10 m height ($h_{ref}$) for the same location, which is normally the reference height of power law wind profile. $\alpha$ is the exponent of power law wind profile, set as 0.3 for a dense urban area [64-66].

The mean radiant temperature ($T_{mrt}$) could be calculated through Equation (5), as a function of global temperature ($T_g$), wind speed ($V_w$) and air temperature ($T_a$) [16], where the global temperature ($T_g$) could be estimated by air temperature ($T_a$), wind speed ($V_w$) and solar radiation $S_0$ by Equation (6):

$$T_{mrt} = [(T_g + 273.15)^4 + 2.47 \times 10^8 \times V_w^{0.6} \times (T_g - T_a)]^{0.25} - 273.15 \qquad (5)$$

$$T_g = T_a + \frac{S_0 - 30}{0.0252S_0 + 10.5V_w + 25.5} \qquad (6)$$

In order to assess the health risk through UTCI values, two thresholds are selected for identifying low (26 °C) and high (38.9 °C) health risk based on findings from recent studies. Pantavou [67] used logistic regression to establish the thresholds above which heat-related symptoms are expected to occur. The study identified 38.9°C (very strong heat stress) as critical limits where the likelihood of heat-related symptoms, such as dizziness and exhaustion, significantly increased during summer time. Di Napoli *et al.* [68] assessed heat-related health risk in Europe using a 38-year dataset, and found that mortality rates rise notably at UTCI values above 26°C, particularly under conditions of moderate and strong heat stress. These findings justify the selection of 26°C and 38.9°C as threshold UTCI values for low and high health risk conditions, respectively.

## 3. Results and Discussions

**3.1 Field measurements of microclimate on four selected sites of NRC campus**

Based on the field measurements, the microclimate data collected at four different locations ('Lawn', 'Parking', 'Tree', and 'Forest') within the NRC campus during the summer of 2024 showed distinct variations in microclimate due to the unique urban features of each location. In Fig. 5, the daily climatic variables measured on the NRC campus are given. The daily air temperature was highest at the 'Parking' location, this site being the one with the least vegetation and having primarily asphalt surfaces. In contrast, the 'Forest' site exhibited the lowest daily air temperature due to the dense tree coverage, which provided consistent shading. The 'Lawn' and 'Tree' locations had moderate temperatures, reflecting the cooling effect from the limited surrounding urban greening, mainly due to the evapotranspiration and shading.

Daily wind speed also varied significantly across the four locations. The 'Lawn' and 'Parking', sites were recognized as open spaces that experienced higher wind speeds as compared to the 'Tree' and 'Forest' sites. Relative humidity was notably higher in the 'Forest' site due to lower solar radiation and higher moisture retention within the forest environment, and it was lowest at the 'Parking' site due to the direct exposure to sunlight and the lack of any urban greening. The 'Tree' and 'Forest' sites experienced significantly lower solar radiation levels due to the shading effect of the local vegetation.

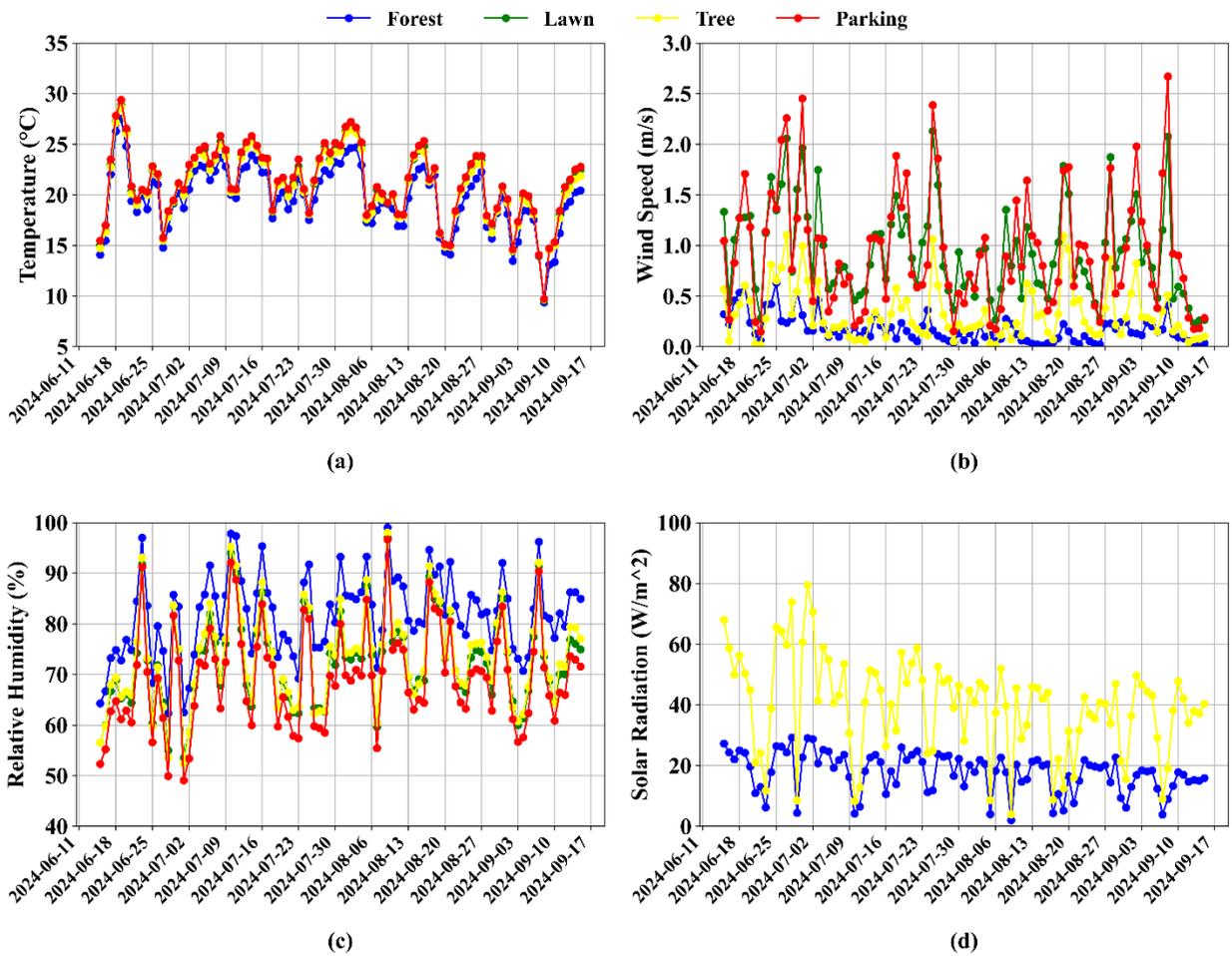

Fig. 5 Daily climatic variables measured on the office campus during 2024 summer time. (a) Daily Air Temperature. (b) Daily Wind Speed. (c) Daily Relative humidity. (d) Daily solar radiation.

In the following Table 3, the values for average and standard deviation of air temperature, wind speed, relative humidity, and solar radiation are summarized for the four sites. Air temperature measurements showed a consistent pattern of higher temperatures in urbanized or exposed areas ('Parking' and 'Lawn') compared to areas having more urban greening ('Tree' and 'Forest'). For instance, the average air temperature in July, 2024, at the 'Parking' site was 23.0 °C, which was 1.7 °C higher than the 'Forest' site (21.3 °C). This difference highlights the cooling effect of vegetation in mitigating the heating effect.

Wind speed data indicated that open areas ('Lawn' and 'Parking') consistently experienced higher average wind speeds. In June, 2024, the 'Lawn' site had an average wind speed of 1.22 m/s, significantly higher than the 'Forest' site with 0.35 m/s, showing the reduction of airflow due to the presence of dense vegetation. Relative humidity (RH) measurements showed that the 'Forest' site maintained the highest RH levels, with an average of 84.8% in August, whereas the 'Parking' area had the lowest RH of 72.4%. This difference of over 12% emphasizes the moisture retention capability of urban greening areas compared to paved surfaces. 'Tree' and 'Forest' sites recorded low values of solar radiation, highlighting the impact of vegetation cover in modulating solar exposure.

Table 3 Field measurements of climatic variables during the summer months on four urban areas

| Location | June | | July | | August | | September | |
|---|---|---|---|---|---|---|---|---|
| | Mean | Std | Mean | Std | Mean | Std | Mean | Std |
| **Temperature** | | | | | | | | |
| Forest | 20 | 4.8 | 21.3 | 3.2 | 19.5 | 3.9 | 16.4 | 4.5 |
| Tree | 20.8 | 5.1 | 22.5 | 3.6 | 20.6 | 4.5 | 17.5 | 5.0 |
| Lawn | 20.9 | 5.2 | 22.6 | 3.9 | 20.8 | 4.8 | 17.7 | 5.4 |
| Parking | 21.3 | 5.2 | 23.0 | 3.9 | 21.0 | 4.8 | 18.0 | 5.4 |
| **Wind speed** | | | | | | | | |
| Forest | 0.35 | 0.30 | 0.15 | 0.19 | 0.11 | 0.14 | 0.13 | 0.16 |
| Tree | 0.50 | 0.57 | 0.28 | 0.41 | 0.31 | 0.41 | 0.27 | 0.40 |
| Lawn | 1.22 | 0.97 | 0.94 | 0.80 | 0.85 | 0.75 | 0.78 | 0.87 |
| Parking | 1.22 | 1.06 | 0.87 | 0.86 | 0.84 | 0.77 | 0.96 | 1.05 |
| **Humidity** | | | | | | | | |
| Forest | 76.4 | 15.3 | 81.9 | 14.2 | 84.8 | 11.8 | 81.0 | 12.3 |
| Tree | 68.9 | 17.2 | 73.5 | 16.7 | 76.4 | 15.4 | 72.4 | 16.0 |
| Lawn | 68.6 | 16.8 | 72.5 | 16.7 | 75.4 | 15.7 | 71.4 | 16.3 |
| Parking | 65.3 | 18.0 | 69.2 | 17.8 | 72.4 | 16.6 | 68.2 | 16.9 |
| **Solar Radiation** | | | | | | | | |
| Forest | 20.2 | 24.7 | 19.7 | 23.7 | 15.9 | 21.6 | 14.6 | 19.0 |
| Tree | 47.3 | 64.6 | 43.8 | 53.6 | 33.6 | 40.7 | 37.0 | 45.1 |

The values of standard deviation (Std) across the different locations provide insights into the variability of the measured climate parameters. For air temperature, the Std was higher at the 'Lawn' and 'Parking' (5.4 °C) sites as compared to the 'Tree' and 'Forest' sites (5.0 °C and 4.5 °C), reflecting larger temperature fluctuations in open areas exposed to direct sunlight. In contrast, the shaded conditions at the 'Tree' and 'Forest' sites resulted in more stable temperatures with lower Std values. In terms of wind speed, the Std was highest at 'Lawn' and 'Parking' sites (0.87 m/s and 1.05 m/s), where open space allowed for more variability in wind conditions. The lowest wind speed variability was observed in the 'Forest' site (0.16 m/s), where the dense tree cover acted as a windbreak, reducing fluctuations in wind speed. Relative humidity also exhibited similar trends, with the lowest Std observed at the 'Forest' site (12.3%) due to consistent moisture retention, whereas the 'Parking' area showed the highest variability (16.9%), reflecting its limited capacity to retain moisture. In summary, air temperature and wind speed were highest in open, urbanized settings such as the 'Parking' and 'Lawn' areas, whereas relative humidity was highest, and variability lowest, in densely vegetated areas such as the 'Forest' site, highlighting the significant microclimatic cooling and moisture retention benefits of urban greening.

### 3.2 Performance of ML methods

The performance of the machine learning models—Artificial Neural Network (ANN), Recurrent Neural Network (RNN), and Long Short-Term Memory (LSTM)—was evaluated based on their ability to predict microclimatic variables (temperature, wind speed, relative humidity, and solar radiation) at the four measurement sites ('Lawn', 'Parking', 'Tree', and 'Forest'). The performance of these models were compared using the Mean Absolute Error (MAE) and $R^2$, as shown in Equation (1) and (2). Table 4 shows the machine learning model performance of climatic variables predictions on four urban areas.

For the prediction of air temperature, it shows the LSTM model outperformed the other ML models, achieving the lowest MAE across all four sites. For instance, at the 'Lawn' site, the LSTM model had an MAE of 0.50 °C, compared to 0.70 °C for the ANN model and 0.71 °C for the RNN model, indicating the ability of the LSTM to better capture temporal dependencies. The $R^2$ value for the LSTM was also consistently higher, reaching 0.98 for all locations, showcasing the model's effectiveness in fitting the data. The ANN model provided a simple and quick baseline for comparison but was less effective in capturing the temporal relationships present in the data.

Table 4 Machine learning model performance of climatic variables predictions

| Station | ANN | RNN | LSTM |
|---------|-----|-----|------|

|         | R²   | MAE  | R²   | MAE  | R²   | MAE  |
|---------|------|------|------|------|------|------|
| **Temperature** ||||||| 
| Lawn    | 0.96 | 0.70 | 0.96 | 0.71 | 0.98 | 0.50 |
| Parking | 0.96 | 0.68 | 0.95 | 0.74 | 0.98 | 0.47 |
| Tree    | 0.96 | 0.67 | 0.96 | 0.67 | 0.98 | 0.4  |
| Forest  | 0.95 | 0.68 | 0.96 | 0.61 | 0.98 | 0.4  |
| **Wind Speed** ||||||| 
| Lawn    | 0.6  | 0.41 | 0.77 | 0.30 | 0.82 | 0.26 |
| Parking | 0.40 | 0.20 | 0.65 | 0.17 | 0.76 | 0.13 |
| Tree    | 0.38 | 0.22 | 0.66 | 0.17 | 0.75 | 0.13 |
| Forest  | 0.05 | 0.12 | 0.62 | 0.09 | 0.70 | 0.08 |
| **Relative Humidity** ||||||| 
| Lawn    | 0.9  | 4.0  | 0.96 | 0.92 | 0.96 | 2.5  |
| Parking | 0.89 | 4.0  | 0.93 | 3.2  | 0.96 | 2.4  |
| Tree    | 0.89 | 4.0  | 0.93 | 3.2  | 0.96 | 2.3  |
| Forest  | 0.82 | 4.4  | 0.90 | 3.1  | 0.95 | 2.2  |
| **Solar Radiation** ||||||| 
| Tree    | 0.62 | 10.2 | 0.74 | 13.9 | 0.75 | 13.0 |
| Forest  | 0.93 | 3.0  | 0.86 | 4.9  | 0.87 | 4.3  |

For the prediction of wind speed, the LSTM again showed superior performance, having the lowest MAE of 0.26 m/s at 'Lawn' site, as compared to values of 0.41 m/s for ANN and 0.30 m/s for RNN, of the respective ML models. The highest R² value of 0.82 was achieved by the LSTM model, indicating its superior predictive capability in capturing wind variations. The dense vegetation of the 'Forest' site, which resulted in lower and more variable wind speeds, made it more challenging for all models; however, the LSTM still maintained a higher R² of 0.7 compared to values of 0.05 for ANN and 0.62 for RNN, of the respective ML models.

For relative humidity, the LSTM model again demonstrated better accuracy with an MAE of 2.2% at the 'Forest' site, compared to 4.4% for ANN and 3.1% for RNN. In terms of the prediction of values for solar radiation, the LSTM model performed well for both the 'Tree' and 'Forest' sites, achieving an MAE of 4.3 W/m² at the 'Forest' location and 13.0 W/m² at the 'Lawn' location. Despite slightly higher MAE in solar radiation, the LSTM performance is much better in predicting the other three climatic variables.

The consistently superior performance of the LSTM model can be attributed to its gated architecture, which enables it to learn and retain long-term dependencies crucial for microclimate prediction [71]. These results resonate with findings from a recent study that utilized a comprehensive three-and-a-half-year dataset to compare LSTM, Random Forest, and Support Vector Regression for solar radiation forecasts [72]. While the study noted that Support Vector Regression performed best for total daily solar radiation, LSTM excelled at hourly and 30-minute predictions, achieving R² values as high as 0.9906 and 0.9956, respectively. This highlights LSTM's adaptability in handling high-resolution temporal data and underscores that large datasets can yield highly accurate predictions, even when certain parameters (such as daily maximum temperature and relative humidity) are excluded. At the 'Forest' site in our analysis—where wind speed fluctuations and dense vegetation introduce both noise and spatial complexity—the LSTM's gating mechanisms likely helped filter out irrelevant temporal information, resulting in lower MAE and higher R² compared to the other models. This capacity to capture nuanced temporal patterns is especially beneficial in microclimatic settings, where conditions can vary rapidly and differ significantly among locations.

The comparison between predicted climatic variables by the LSTM model and observed values from all locations, as shown in Fig. 6, further demonstrates the better performance of the LSTM among the three models investigated in this paper. The scatter plots indicate a strong agreement between predicted and observed values for all climatic variables, with points clustering closely around the $y = x$ line. The LSTM model effectively captured the dynamics of air temperature, relative humidity, solar radiation, and wind speed, with most predicted values aligning well with observations, despite slight underestimations at higher wind speeds and some scatter for solar radiation. These visual comparisons highlight the LSTM model's ability to provide reliable and accurate predictions for complex microclimatic conditions.

A direct comparison with Yang *et al.* [37]'s results shows that for air temperature prediction, LSTM model of this study achieves an MAE as low as 0.50 °C and an $R^2$ of 0.98, whereas MLP approach in the previous study reports an MAE of 0.52 °C but attains a slightly higher $R^2$ of 0.996. For wind speed, LSTM attains an $R^2$ of up to 0.82 and an MAE of 0.26 m/s, outperforming ANN method of previous study, which yielded an $R^2$ of 0.50–0.67 and an MAE of approximately 0.52–0.87 m/s. These numerical comparisons suggest that LSTM-based framework provides competitive temperature predictions while delivering notably stronger performance in forecasting wind speed. Moreover, the present study extends beyond temperature and wind speed predictions to encompass additional variables such as relative humidity and solar radiation, neither of which were addressed in previous work.

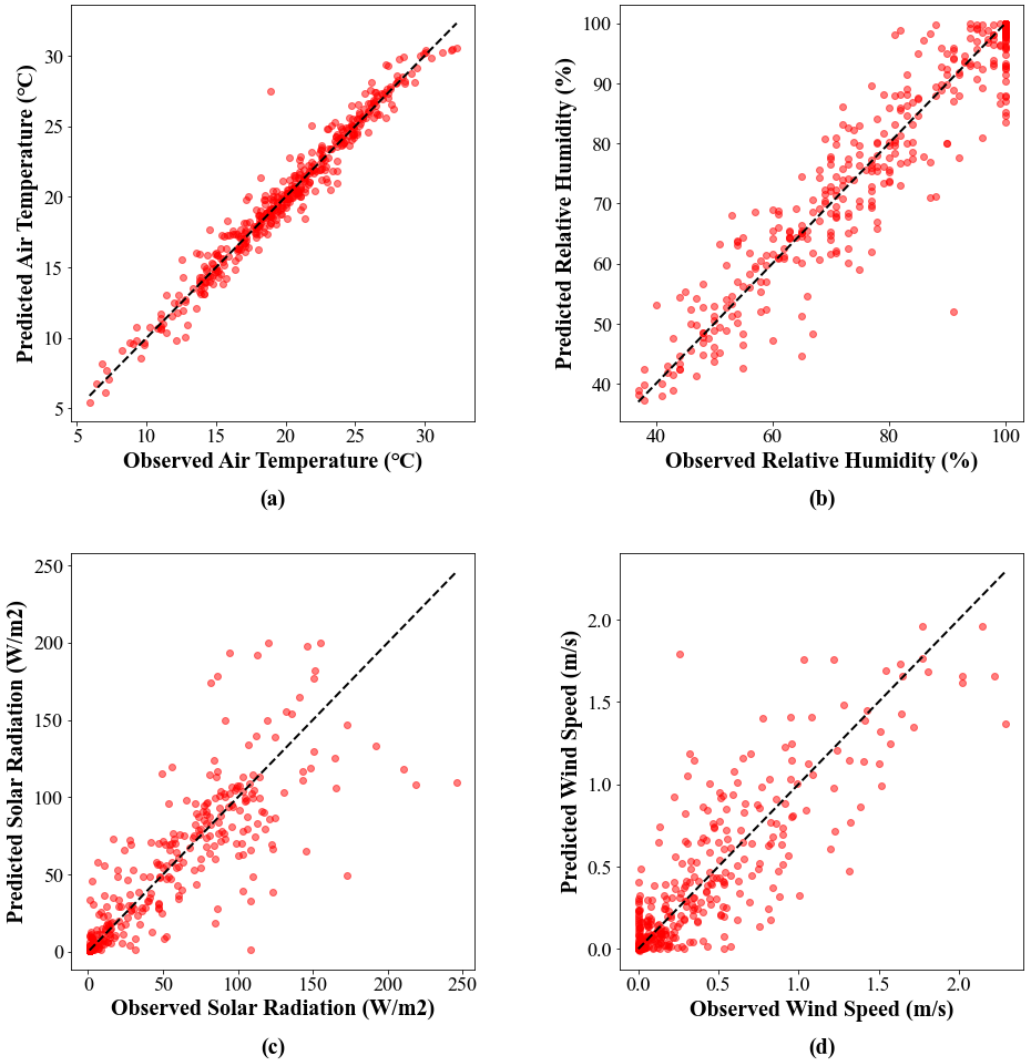

Fig. 6 Comparison between predicted climatic variables by LSTM (y-axis) and observations (x-axis)

Overall, the results indicate that, whereas all three machine learning models provided valuable insights, the LSTM model consistently outperformed the ANN and RNN models, particularly for capturing the temporal dependencies inherent in microclimatic data. The LSTM's memory cell structure allowed it to more effectively account for long-term dependencies, which was crucial for predicting conditions that exhibit significant temporal variations, such as: wind speed, relative humidity, and solar radiation. Thus, in this study, the LSTM was selected for future climatic variable predictions.

**3.3 Impacts of Climate Change and Urban Greening on Urban Overheating**

The LSTM model, selected from the above section, was used to predict future climatic variables of four (4) locations and for calculating future local UTCI values. The impact of climate change on the value of the UTCI at all locations is evident across different GW scenarios, as illustrated in Section 2.3. In Fig. 7, the daily average value for the UTCI, as predicted, is reported for four locations of the NRC campus under various GW scenarios during the months of July and August, 2024. Under GW1.0, the UTCI value at the 'Parking' site averaged ca. 27 °C during the summer period. As global warming scenarios intensified, the UTCI value also increased: under GW1.5, it reached approximately 28 °C; under GW2.0, it rose to ca. 29 °C; and under GW3.5, it further increased to approximately 31 °C.

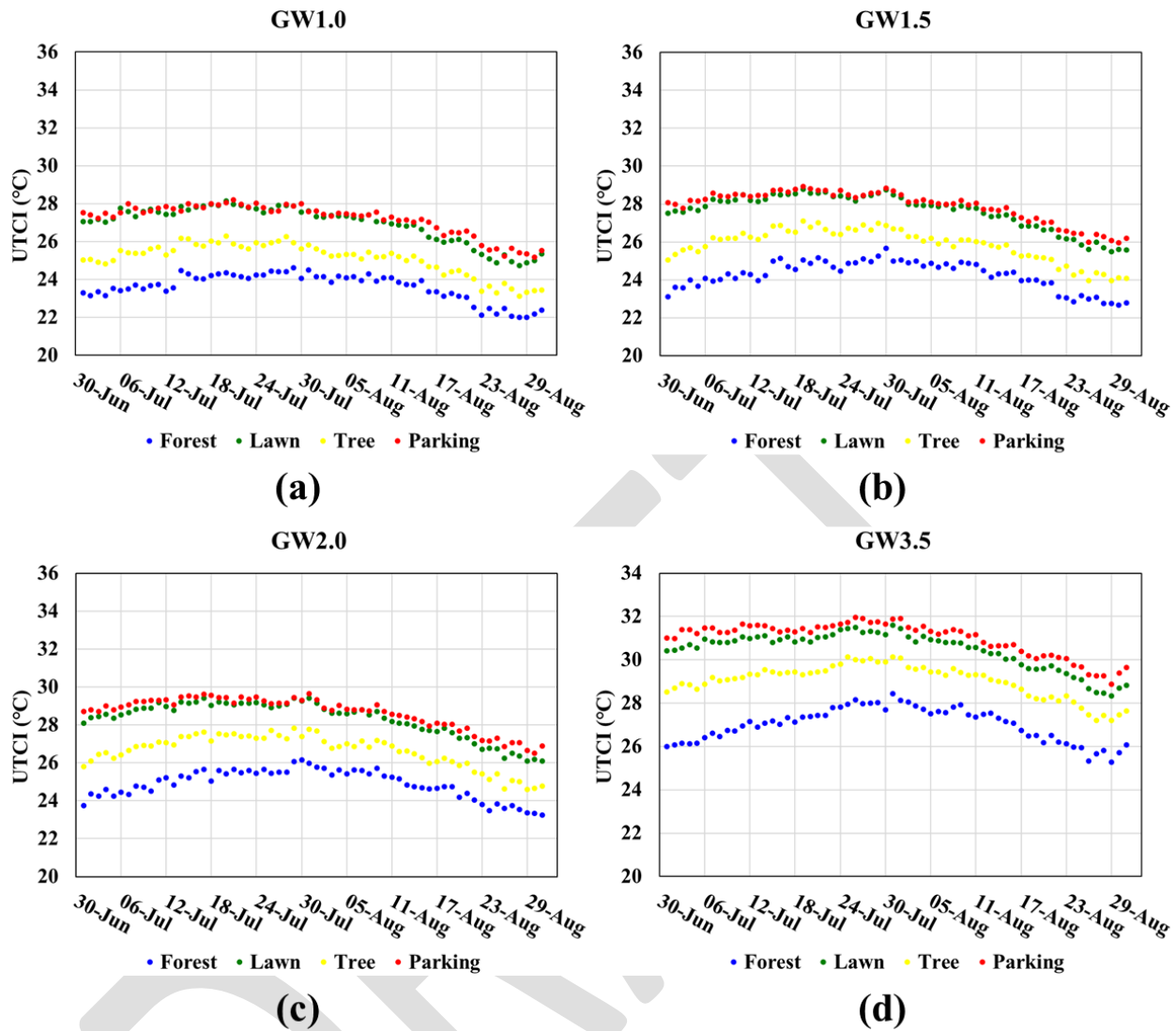

Fig. 7 Daily average UTCI value prediction for four locations on the NRC campus under (a) GW1.0, (b) GW1.5, (c) GW2.0, and (d) GW3.5, for the months of July and August, 2024

Urban greening, as represented by the 'Lawn', 'Tree', and 'Forest' sites, has a significant mitigating effect on air temperature compared to the 'Parking' location under different GW scenarios. The 'Lawn' site, with less greening, showed a slight reduction in UTCI compared to the 'Parking' site across all scenarios. Under GW1.0 and 1.5, the 'Lawn' site shared close UTCI as 'Parking', with difference less than 0.5 °C. Under GW2.0 and GW3.5, the 'Lawn' site recorded UTCI values of

28 °C and 30 °C, respectively, consistently maintaining temperatures 0.5-1 °C lower than the 'Parking' site. The 'Tree' site provided better cooling effects. Under GW1.0, the UTCI at the 'Tree' site averaged around 25 °C, resulting in a 2 °C reduction compared to 'Parking'. This trend continued under GW1.5, GW2.0, and GW3.5, with UTCI values of 26 °C, 26 °C, and 29 °C, respectively, indicating a consistent reduction of 2-3 °C compared to the 'Parking' location. The presence of trees significantly contributed to shading and evapotranspiration, which helped to limit the temperature increases.

The 'Forest' site demonstrated the most substantial cooling effect. Under GW1.0 and 1.5, the UTCI at the 'Forest' site averaged around 24 °C, which was 4 °C lower than the 'Parking' site. Under GW2.0 and 3.5, the UTCI increased to 25-27 °C, providing a reduction of 4-5 °C compared to the 'Parking' location. The dense vegetation in the 'Forest' site provided effective shading and cooling through evapotranspiration, significantly mitigating the effects of rising temperatures under changing climate. These comparisons clearly demonstrate that urban greening, whether through lawns, the presence of trees, or dense forests, plays a crucial role in reducing urban overheating risk and enhancing thermal comfort. The effectiveness of urban greening in mitigating temperature rise becomes even more critical under intensified global warming scenarios, underscoring the importance of incorporating green infrastructure into urban planning to combat the effects of climate change.

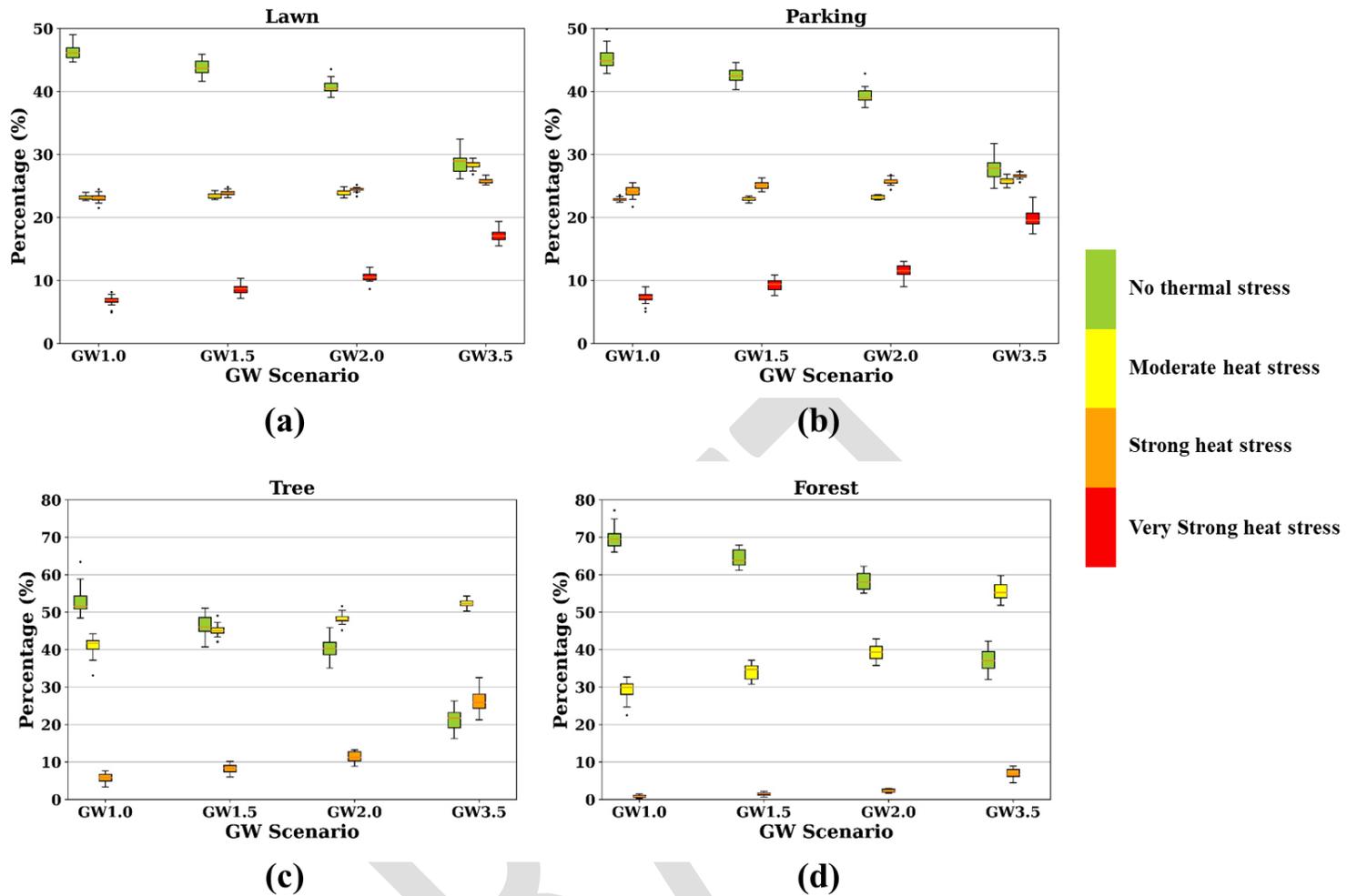

Fig. 8 Prediction of the percentage of thermal stress among two summer months considering the uncertainty from 15 runs under four GW scenarios.

The prediction of thermal stress under different global warming scenarios reveals significant differences in how various greening strategies influence thermal comfort. The boxplots in Fig. 8 illustrate the percentage of time each location ('Lawn', 'Parking', 'Tree', and 'Forest') experienced different levels of thermal stress (No Thermal Stress, Moderate Heat Stress, Strong Heat Stress, and Very Strong Heat Stress) during the summer months, considering the uncertainty from 15 model runs under each GW scenario. By 2050, if SSP2-4.5 occurred (corresponding to GW1.5), the UTCI value at the 'Parking' location would indicate an increase in strong and very strong heat stress, in total, reaching these values nearly 35% of the time. By 2090, if SSP5-8.5 occurred (corresponding to GW3.5), very strong heat stress would occur at the 'Parking' site almost 20% of the time and total overall time of strong and very strong heat stress would increase to 46%. This clearly highlights the vulnerability of non-vegetated areas to achieve elevated thermal stress levels under more intense global warming scenarios.

Whereas the 'Tree' and 'Forest' sites exhibited greater uncertainty in thermal stress predictions, they nonetheless, consistently experienced lower UTCI values as compared to the 'Lawn' and 'Parking' sites. For the 'Tree' site, the main increase in thermal stress was observed in strong heat stress, rising from 5% under GW1.0 to 25% under GW3.5. In contrast, the 'Forest' site primarily experienced a large increase in moderate heat stress, from 29% under GW1.0 to 55% under GW3.5, indicating that the dense vegetation was effective in mitigating more severe stress levels. In comparison, 'Lawn' and 'Parking' locations experienced an increase primarily in very strong heat stress. At these sites, very strong heat stress increased from around 7% under GW1.0 to around 19% under GW3.5, further emphasizing the importance of green infrastructure. Overall, the findings demonstrate that urban greening plays a crucial role in reducing not only the intensity but also the frequency of severe thermal stress, thereby enhancing thermal comfort under future climate scenarios.

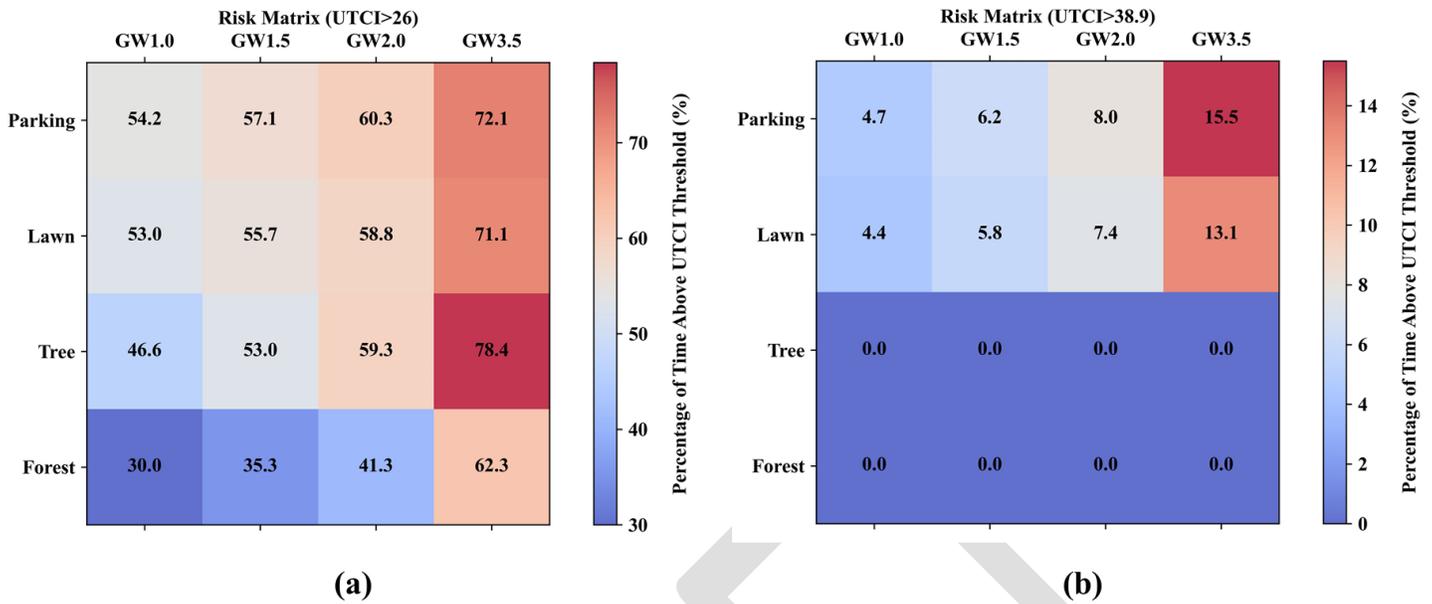

Fig. 9 Percentage of time of high and low health risk conditions at different locations under various GW scenarios based on UTCI values (low risk: UTCI > 29 °C, high risk: UTCI > 38.9 °C)

The health risk assessment matrix, depicted in Fig. 9, illustrates the percentage of time of low and high health risk conditions at different locations, the low and high risk was determined by different UTCI thresholds under the four global warming (GW) scenarios. The left figure represents the percentage of time above a UTCI of 26°C (low health risk), while the right figure represents the percentage of time above a UTCI of 38.9°C (high health risk), based on Section 2.4.

For low health risk (UTCI > 26°C), the results show a significant increase in the percentage of time across all locations as the GW scenarios intensified. At the 'Parking' and 'Lawn' sites, the percentage of time above the low risk threshold increased from 54.2% and 53% under GW1.0 to 72.1% and 71.1% under GW3.5, respectively. The 'Tree' and 'Forest' sites also experienced an increase. The 'Tree' site showed an increase from 46.6% under GW1.0 to 78.4% under GW3.5, while the 'Forest' site had the lowest increase, rising from 30% to 62.3%. It shows that the low health risk will be increased in all locations under the climate change impacts, regardless of the urban greening conditions.

For high health risk (UTCI > 38.9°C), the results indicate that both the 'Parking' and 'Lawn' sites experienced a notable increase in the percentage of time under health-threatening conditions with the GW levels increased. Under GW1.0, the percentage of time above this threshold was 4.7% for 'Parking' and 4.4% for 'Lawn'. For GW3.5 scenario, these values increased to 15.5% for 'Parking' and 13.1% for 'Lawn'. In contrast, the 'Tree' and 'Forest' sites did not experience any time above this high risk threshold, even under the highest GW scenario. This stark difference emphasizes the role of urban greening in eliminating the occurrence of extremely high-risk heat conditions, suggesting that dense vegetation like trees and forests is crucial for maintaining safer thermal environments in the face of escalating climate change.

## 4. Conclusions

The study presented significant contributions to understanding the potential of urban greening as a mitigation strategy for urban overheating under future climate change scenarios. By integrating local field measurements with machine learning models, this research was useful in predicting future urban overheating at local sites with limited measurement data under various urban greening settings. The main findings of this study are the following:

- Based on the field measurements, air temperature and wind speed were highest in open urbanized areas such as the 'Parking' and 'Lawn' areas, whereas relative humidity was highest and variability lowest in densely vegetated areas such as the 'Forest' area, this highlighting the benefits of urban greening to bring about significant microclimatic cooling and moisture retention;

- The LSTM model consistently outperformed the ANN and RNN models. It could cope with long-term dependencies more effectively, which was crucial for predicting conditions that exhibit significant temporal variations, such as: wind speed, relative humidity, and solar radiation,.

- As global warming scenarios intensified, UTCI at the 'Parking' site increased from around 27 °C under GW1.0 to 28 °C under GW1.5, 29 °C under GW2.0, and approximately 31 °C under GW3.5.

- By 2050, under SSP2-4.5 (GW1.5), the 'Parking' site could experience strong and very strong heat stress 35% of the time, increasing to 46% by 2090 under SSP5-8.5 (GW3.5), thereby highlighting the vulnerability of non-vegetated areas to extreme thermal stress as global warming intensifies.

- Low health risk (UTCI > 26°C) will be increased in all locations under the impact of climate change, regardless of urban greening conditions.

- Urban areas having dense vegetation such as 'Tree' and 'Forest' areas, are effective in eliminating the occurrence of extremely high-risk heat conditions (UTCI > 38.9°C).

There are several limitations to this study. Firstly, the field measurements were conducted over only a single summer period, thus limiting the reliability of the predicted future climate data due to the restricted range of climatic variables captured within the three month period. Additionally, whereas various urban settings were considered, the degree to which of the conclusions of thid study can be generalized—particularly the reduction in UTCI and thermal stress through urban greening—remains uncertain. Further studies are needed to replicate the findings across more urban settings having similar greening conditions. Lastly, despite advancements in machine learning algorithms, numerous newly developed methods may outperform LSTM. However, due to the limited availability of public code and resources, only three machine learning models and a selected LSTM were evaluated in this study; this work, nonetheless initiates research in this field of study.

## Acknowledgement

The authors are grateful for financial support from Building Envelope and Material Resource Unit in Construction Research Centre at National Research Council Canada, "Nature Based Solutions to Mitigate Urban Heat Islanding Effects of Canadian Communities" [#1019534], the Natural Sciences and Engineering Research Council (NSERC) of Canada through the Discovery Grants Program [#RGPIN-2024-06297], and the SEED project Creating Electrified and Decar-bonized Healthy Urban Microclimate around Building Clusters through Climate-Resilient Solutions under the Canada First Research Excellence Fund (Volt-Age). We would like to thank Chantal Arsenault for sharing measured climate data, James Adam, Gnanamurugan Ganapathy, Josip Cingel, and Mike Nicholls for their invaluable assistance in building the weather stations and supporting fieldwork efforts.